# The Unbearable Lightness of Prompting: A Critical Reflection on the Environmental Impact of genAI use in Design Education

**Maria Luce Lupetti, Elena Cavallin, and Dave Murray-Rust**

**ABSTRACT**
Design educators are finding ways to support students in skillfully using Generative Artificial Intelligence (GenAI) tools in their practices while encouraging the critical scrutiny of ethical and social issues around these technologies. However, the problem of environmental sustainability remains largely unaddressed. There is a lack of both resources to grasp the environmental costs of genAI in education and a lack of shared practices around the issue. This work contributes filling this gap by counting the energy costs of using genAI in design education and critically reflecting on the impact of these costs. We leverage the image data collected during a genAI workshop for designers held in 2023 with 49 students, to calculate the energy costs of these types of activities. The results reveal that a genAI workshop for designers can easily double the energy costs associated with students' use of computers, countering the efforts of educational institutions to minimize their energy expenditure. We critically reflect on this finding to distill a set of five alternative stances, with related actions, that can support a conscious use of genAI in design education, while respecting individual positions. The work contributes to the field of design pedagogy, and education more broadly, by bringing together ways for educators to reflect on their practices and informing the future development of educational programs around genAI.

## 1. INTRODUCTION

Artificial Intelligence (AI), and more specifically Generative Artificial Intelligence (genAI), have recently seen large adoption in a great variety of sectors (Chan and Colloton, 2024). In particular, genAI is now being extensively used by design professionals and researchers across all phases of the design process (van der Maden et al., 2024), from the `fuzzy front end' (Sanders and Stappers, 2008) of ideation, content generation, and quick production of mood boards and product visualizations, up to enabling fast and cheap product evaluation with synthetic users (Li, 2024). In the face of these technological possibilities, academics have started to study emerging human-AI design practices (Flechtner and Stankowski, 2023; Subramonyam et al, 2022; Audry, 2021; Olsson and Väänänen, 2021; Fu and Zhou, 2020; Fiebrink, 2019; Yang et al., 2018; Dove et al., 2017), and question how to re-think their educational programs (Tien and Chen, 2024; Whitham et al., 2024; Huang et al., 2023). As a result, the last two years have seen a proliferation of workshops, elective courses, exercises and other initiatives to confront design students with the opportunities and challenges brought about by genAI.

This increasing engagement of design educators with genAI is often driven by the belief that genAI is a powerful innovation to be embraced, as it can make us more creative (Lim et al., 2023; Figoli et al., 2022). However, engagement with genAI can take a more reactionary stance, as a way to adjust to a seemingly inevitable technological future (van der Maden et al., 2024; Bozkurt, 2023; Fathoni, 2023), where the designers are seen as playing a strategic role in protecting the interests and values of users (Armstrong, 2021). The inherently complex and controversial nature of genAI imposes on educators a responsibility to nurture a critical sensitivity in future professionals so that they can grasp the underlying mechanisms driving these systems while consciously engaging with their ethical and social implications (Chan and Colloton, 2024), ultimately aiming to ensure that the new generations will shape technologies that are respectful of human and collective values (Tubella et al., 2023). Thus, alongside training students to skillfully handle genAI as design material, with its very capabilities and limitations (Casal-Otero et al., 2023), educators are now developing novel approaches for training designers to understand, anticipate and, if necessary, mitigate the effects of genAI technologies in society (Huang et al., 2023; Murray-

Rust et al., 2023; McDonald et al., 2021). These initiatives focus on a range problems, including algorithmic biases and related consequences of social injustice (Aguilar, 2024; Murray-Rust et al., 2023; McDonald et al., 2021) as well as developing disciplinary knowledge about the possible relationships and dynamics between design and AI (Lupetti and Murray-Rust, 2024; Murray-Rust et al., 2023; Figoli et al., 2022; Simeone et al., 2022; Hemment et al., 2024; Liu et al., 2024b; Tsiakas and Murray-Rust, 2024; Murray-Rust et al., 2024). In spite of all this, a key concern around genAI in design education is left unspoken: the one of *environmental sustainability*.

The sustainability problem of genAI has received increasing attention in academic communities, with experts pointing at the massive costs that genAI carries in terms of water (Li et al., 2023), electricity (Luccioni et al., 2024), and $CO_2$ (Wu et al., 2022). Despite making headlines in popular newspapers (i.e., The Guardian (Stokel-Walker, 2023), and Forbes (Zvibel, 2024) just to name a few) and being a central argument in keynote talks at important academic events (see the CHI keynote talk by Crawford (2024a)), however, the *environmental impacts of genAI remain predominantly unaddressed in design education*. The argument put forward in this work is that designers, and design educators more specifically, still lack resources and practices for grasping the environmental impact of their educational practices when engaging with genAI. And although, there is a growing push within the field of design for education to be made sustainable (e.g. (Faludi et al., 2023; Faludi et al., 2019), actions are still lacking to bridge these efforts with the ones aimed at nurturing AI literacy. Even looking beyond the design field the problem seems to remain largely unspoken. Even the UNESCO's report on *Guidance for generative AI in education and research* (Holmes et al., 2023), a seminal resource in the space of AI and education, contains only one small recommendation in this direction, that is to "Analyse the environmental costs of leveraging AI technologies at scale (e.g. the energy and resources required for training GPT models), and develop sustainable targets to be met by AI providers in a bid to avoid adding to climate change" (Holmes et al., 2023). But, *how should educators practically do that?*

There is no established method or examples that inform the design education community on how to calculate the environmental costs of running a workshop or an exercise involving genAI, or how to analyze the costs in context. This leaves educators to make their own decisions on whether to engage with these technologies or not primarily grounding their choices on gut feelings and personal beliefs. This work, then, aims to provide design educators with an example of *how to perform a research-grounded sustainability accounting of design educational programs engaging with genAI*, and provides actionable recommendations for taking a stand in this space. We first draw a picture of the current scientific critique around the problem of environmental sustainability associated with AI. From this, we extract data about the energy consumption required by AI-generated images and use this value to calculate the energy impact of a genAI workshop for designers. We account for the environmental impact of the images produced in a two-day workshop designed and conducted by the first author in collaboration with the second author. We draw on the results of this calculation together with AI sustainability literature to reflect on the impact of these practices and to derive a set of *stances* illustrating alternative positions that design educators may identify with, along with related *actions* they can take to reduce the environmental impact of genAI use in design education. The paper concludes with a discussion of the practical challenges associated with the two contributions of this work: a *blueprint for counting* the energy costs of design education engaging with genAI, and a disciplinary *call for sustainability accounting* in design universities.

## 2. AI IN DESIGN EDUCATION

> *"Given that many of my students are already using these tools and their use will likely be prevalent in the industry in the near term, I feel I should ask: to what extent should I incorporate generative AI into my pedagogies and curricula?"* (York, 2023)

Many design educators have integrated AI tools and techniques into their programs and curricula in recent years, perhaps sharing a position similar to York (2023), cited above. In the face of a rapidly changing technological landscape, in which every month there is a new release of genAI tools of some sort, design educators find themselves wondering what new skills and knowledge will be required of the next generation of designers (Whitham et al., 2024; Pei et al., 2023; York, 2023) and how to incorporate these tools in existing programs (Maceli et al., 2024; Flechtner and Stankowski, 2023; Goel et al., 2023) -- not to mention the vexing question of how to keep up with all this themselves (Flechtner and Stankowski, 2023).

There is a history of design tools that engage with 'AI', many of which date from before the explosion of genAI. Tools like Wekinator (Fiebrink et al., 2010), Teachable Machine (Carney et al., 2020) and Edge Impulse (Banbury et al., 2023) have been widely adopted in education, where they allowed educators to reduce the barrier to entry of AI and machine learning (ML) and to make these more accessible for designers and creative practitioners (Flechtner and Kilian, 2024). These experiences contributed raising an awareness in the field that creative practitioners need distinct knowledge, tools, and examples and that it is essential to make space for experiential learning where students "iterate through a cycle of concrete experience, reflective observation, abstract conceptualization, and active experimentation" (Fiebrink, 2019).

Sharing similar views on the importance of experiential learning and meaningful 'translations' of AI concepts, several educators have recently engaged ways in which designers can relate to AI systems. Some explore the use of metaphors in making sense of and using AI (Murray-Rust et al., 2024; Tian et al., 2024; Murray-Rust et al., 2022; Dove and Fayard, 2020), others look at useful framings of AI technologies, i.e., not as mere tools but rather partners (Figoli et al., 2022; Simeone et al., 2022), up to the point of rethinking interactions as co-performances \cite{kuijer2018co}. Alongside these theoretical works, many practical resources for designers and design educators have also been developed in recent years. For instance, the online platform *Designing With AI* collects resources and methods for designers working with AI (Botta et al., 2024). The *Mix & Match* toolkit supports designers' ideation processes around ML (Jansen and Colombo, 2023). And the *Experiential AI Exercises* by Murray-Rust and colleagues (Murray-Rust et al., 2023) consist of quick activities that facilitate students' engagement with AI interactional affordances, matters of relationality and wider implications. These are just a few of a multitude of design education resources that along with introducing students to the technical functioning of AI, also share the commitment --more or less explicit-- to nurturing *AI literacy* (Chan and Colloton, 2024), i.e., to equip future generations with both practical understanding of genAI capabilities and limitations (Casal-Otero et al., 2023), but also with a critical capacity to understand, anticipate and (if necessary) mitigate the effects of these technologies in society (Murray-Rust et al., 2023; McDonald et al., 2021).

### 2.1. Critical perspectives in AI education for designers

The more genAI is integrated into educational settings, the more educators raise concerns over the consequences of this transition. Concerns range from purely didactic apprehensions about matters of originality, plagiarism, and learning outcomes (Whitham et al., 2024; Sandhaus et al., 2024; Lim et al., 2023; Fathoni, 2023; Oravec, 2023), to the question of how to support students understanding the implications of genAI at societal scales (Chan and Colloton, 2024; Vartiainen and Tedre, 2023). Programs are being developed around nurturing AI literacy where the emphasis is not purely on the technical functioning, but rather on possible societal consequences and power relations these can bring (Sandhaus et al., 2024; Aguilar, 2024; Vartiainen and Tedre, 2023; Arada et al., 2023; McDonald et al., 2021). This goes hand-in-hand with a growing collective awareness about the risks and consequences of genAI capabilities when integrated into services and products, that can lead to discrimination and other forms of injustice (Costanza, 2018). Bias and injustice are non-trivial problems, and as Kharrufa and Johnson (2024) illustrate, design students are often contradictorily found to acknowledge the problem of bias, but at the same time fall into the trap of using the perceived objectivity of genAI as a design evaluation tool in place of actual users. Thus, AI literacy programs for designers increasingly also integrate performative, critical and speculative design practices (Rahm, 2024; Costello et al., 2024; Thrall et al., 2024; Flechtner and Kilian, 2024; Murray-Rust et al., 2023) as a way to strengthen critical understanding of AI (Costello et al., 2024) and address its political, economic and ethical implications (Rahm, 2024). These approaches are driven by the belief that designers have an important role to play, in between driving innovation around disruptive technologies and advocating for people's needs (Pei et al., 2023), even when not intimately familiar with the technical functioning of AI (Murray-Rust et al., 2023). Despite their importance and growing presence, however, these critical pedagogic practices leave unattended one of the most essential concerns that have recently interested both the academic and the public discourse around AI: *the problem of environmental sustainability*.

### 2.2. The AI sustainability critique

> "*We need pragmatic actions to limit AI's ecological impacts now*" (Crawford, 2024b)

An awareness is growing, both in the academic community and the public, that AI's impact on society is not solely socio-technical, but ecological too (Rakova and Dobbe, 2023). OpenAI's chief executive Sam Altman recently admitted that the AI industry is heading towards an energy crisis as it requires such large amounts of energy that current energy systems will struggle to cope (Crawford, 2024b; Saul et al., 2024). It has been estimated that a *search driven by genAI uses four to five times the energy of a conventional web search*, and that ChatGPT consumes the same energy as 33000 homes (Crawford, 2024b). Although exact data is hard to get –due to the environmental costs of genAI being treated as secret by the industry (Crawford, 2024b) – recent reports warn about the increasing demand for computational power, and thus data centers, that AI is driving (Yao, 2024; Goldman Sachs, 2024). Data centers already account for 2% of global electricity usage (Li et al., 2023; Patterson et al., 2022), and this is expected to grow dramatically with the uptake of genAI. A study by Bloomberg Technology (Saul et al., 2024) estimates that by 2034, global energy consumption by data centers will match the one currently used by the entire India. And, although these are largely estimations, the constant building and use of new data centers is a fact (Saul et al., 2024; Chen, 2024).

The environmental impact of running AI systems extends beyond electricity use and includes the emission of CO2 and the consumption of fresh water. Further, Li and colleagues (Li et al., 2023) investigated the consumption of clean freshwater attributable to data centers and thus also intertwined with the rising demand for computational power by AI. The authors report that the training of GPT-3 in a state-of-the-art US data center can consume about 700.000 liters of clean freshwater, which would be enough to produce *320 Tesla electric vehicles*, and that a simple conversation with ChatGPT (roughly 20-50 questions and answers) can require a *half liter bottle of water* (Li et al., 2023). These costs can be highly variable -- Dodge and colleagues (Dodge et al., 2022), for instance, found that the training of a language model emits between 10Kt CO2 and 28Kt CO2, depending on where the training happens. Casting sustainability questions in terms of tons of CO2, and liters of fresh water is crucial, yet understanding the comparative impacts of GenAI within an educational curriculum remains hard to grasp, as these impacts also happen for other activities. Parallels can be drawn with concepts like *carbon numeracy* – "the ability to correctly understand and manage one's own carbon footprint" (Wynes, 2020) – that tells us about how difficult it is for people to properly underestimate the environmental impact of our practices, such as eating meat (Camilleri et al, 2019) or car driving (Grinstein et al., 2018). One way to approach this is to provide graspable metrics to communicate meaningfully about the sustainability of actions to allow consumer comparison, such as 'light bulb minutes' (Camilleri et al, 2019), traffic light labels for food (Thorndike et al., 2014) and comparisons such as "reducing your meat intake to three times per week is equivalent to avoiding six short-haul return flights each year" (Steinitz et al., 2024). In the same way, Luccioni and colleagues (2024a) argue that "*giving the public a simple way to make informed decisions would bridge the divide that now exists between the developers and the users of AI models, and could eventually prove to be a game changer*".

## 3. A CRITICAL REFLECTION ON GENAI SUSTAINABILITY IN DESIGN EDUCATION

We share with (Luccioni et al., 2024a; Crawford, 2024b) the view that pragmatic actions are needed to limit the environmental impact of our practices around the use of genAI, and that dedicated actions should facilitate the making of informed decisions in this space. Yet, we see two key difficulties in this path.

Firstly, there is a fundamental problem hindering the way to a proper grasping of the impact resulting from AI-related activities, and it concerns the accessibility and interpretability of data around the problem. The lack of transparency around the environmental costs of genAI (Crawford, 2024b) is often worsened by the encouragement of the misbelief that *'the training is the problem'* (Luccioni et al., 2024b). Even well-meaning initiatives tend to predominantly emphasize the environmental problem of AI in its training phase, leaving the impact that is actually generated at the inference stage largely unattended. While providing correct info, the narrow focus of these critiques prevents us from thinking systemically about AI sustainability, and guides us to focus on the relatively small costs of training, rather than the large –and growing– costs of inference or querying. In fact, the deployment costs of users querying a model "even assuming a single query per user, which is rarely the case, [...] would surpass its training costs after a few weeks or months of deployment" (Luccioni et al., 2024b). Furthermore, it is crucial for educators to understand the impact of genAI querying as, while we can't make a change to the costs associated with training, we can and do have an impact on the costs associated with genAI inferencing and querying, as this is what gets used in education. Recently we have started to witness a growing attention towards the use phase of genAI, yet resources in this space are still scarce. To our knowledge, the benchmark from (Luccioni et al., 2024b) is the only resource available today that clearly provides us with energy and CO2 costs associated with genAI activities. Furthermore, even when provided with this data, it remains hard for educators to translate it into an energy use estimation of their activities.

The second difficulty we see in making informed decisions about genAI use in design education, then, is the problem of what actions to take in response to the available knowledge. Despite a 'turn to sustainability' we can currently find in design education (Faludi et al., 2023), work that explicitly engages with questions of whether we should limit or prohibit the use of genAI in light of the sustainability problem is missing. The only exception (to our knowledge) is the work by Maceli and colleagues (Maceli et al., 2024) which articulates the problem space and provides a clear position. According to the authors, it is important to inform students about the environmental costs of genAI but it is also crucial to prepare them for their future careers. Hence as educators, we should not prohibit them from using these tools, but rather provide them with a critical understanding that hopefully will help them make ethical choices in the future. We see this work as an important step towards reflexivity we should all practice as design educators engaging with genAI. Yet, even in this work the sustainability concern is presented as a reflection point from the authors, but no practical recommendations are provided to neither reduce the impact of our educational practices nor help the community take a stand.

Thereafter, in this paper we provide: i) *a research-grounded approach to counting the energy costs of genAI educational activities*; and ii) *a spectrum of stances and related activities that can help design educators and students in accounting for sustainability when using genAI*. The work and arguments presented here are the results of a critical reflection that stems from the personal teaching experience of the three authors who, over the last four years, integrated machine learning tools and genAI into their education, applying various formats and adjusting to diverse contexts, such as quarter or semester-long courses, workshops of either one or a few days, and online sessions of a few hours. Among these, we decided to leverage the visual materials produced during a two-day workshop for bachelor design students, run at the end of 2023 at *University of San Marino*, to calculate the environmental costs of these activities in terms of electricity use. It has to be noted that the workshop was run before this critical engagement with the theme of AI sustainability, thus, it is not intended to present a methodology for reducing the environmental impact of these activities, but rather to provide a ground for situating the problem of AI sustainability in a specific educational practice we could perform a detailed accounting of. Through critical reflexivity (Mao et al., 2016), that took the form of online discussion sessions, we attempted to move from our personal didactic experiences and their estimated energy costs to envisioning possible practices for consciously embedding genAI in design education. In particular, the lens of *critical reflexivity* helps us frame these practices not as prescriptive actions to take, but rather as alternative possibilities to consider and choose, based on one's personal situations and ideologies.

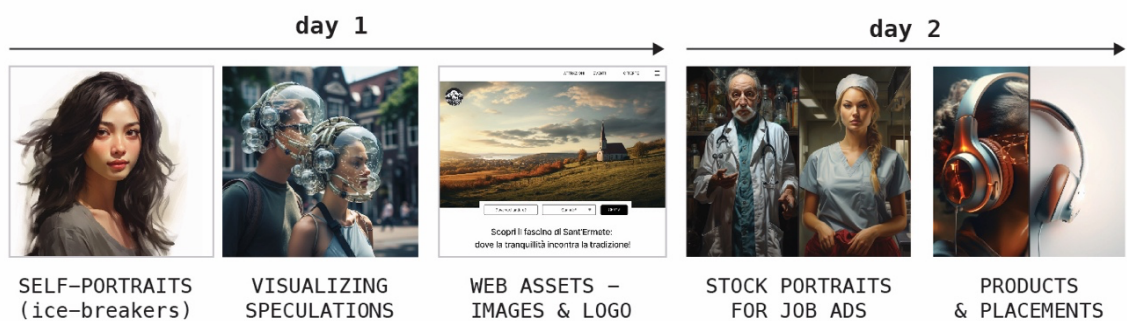

**Figure 1. Overview of the activities conducted during the workshop.**

## 4. COUNTING THE ENERGY COSTS OF A GENAI WORKSHOP

The selected workshop was run over two days, from November 30th to December 1st 2023, *San Marino*, at the *University of San Marino*. The workshop lasted 12 hours, with a full-day program on the first day (8 hours), and a half-day on the second (4 hours). The course was offered as an addition to the main courses offered as part of a bachelor program in Design and attended by a total of 49 students.

The main focus of this workshop was to introduce students to text-to-image genAI tools and get them familiar with ways in which these resources can be used for design activities. As per the explicit request of the local didactic coordinator, the workshop used *Midjourney* as genAI platform, which was selected for the good usability of the platform and the high aesthetic quality of the generated results. The workshop started with an introductory lecture and was followed by a series of exercises in which students had to generate images for different purposes (Figure 1), such as creating website hero images, logos, portrait photos for social media campaigns, visualizing future scenarios, and more. We used *FigJam*, an online whiteboard tool, to prepare templates for each activity, to keep track of students' work, and to facilitate collective sharing.

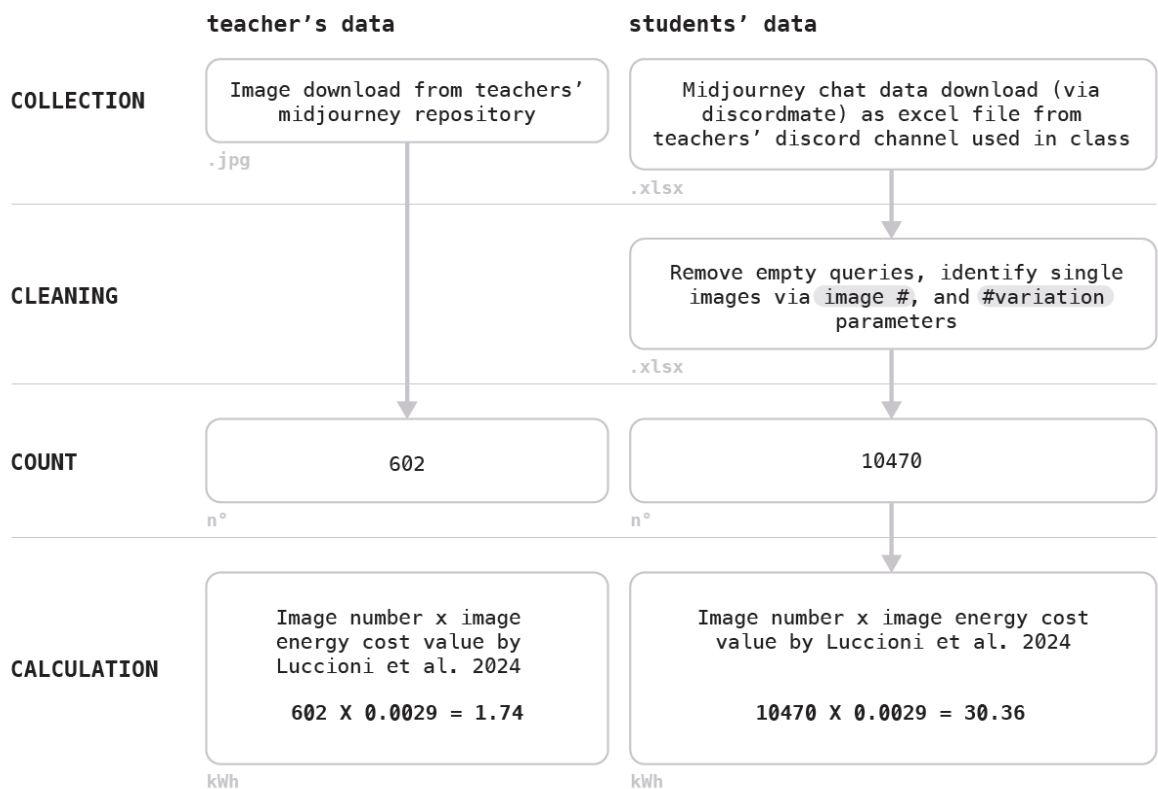

Figure 2. Overview of the activities conducted during the workshop.

### 4.1. Image collection and counting

Here we consider 'data' the images produced with *Midjourney* during the workshop, that were collected and analyzed for calculating the electricity consumption of the activities. The data collection and analysis process, illustrated in Figure 2, started with a collection and categorization of the image data first into two main groups, one of the *teacher* and one of the *students*. Next, we further divided images generated by the teacher into two subgroups: *preparation*, all the images generated before the actual workshop that were used to compile the lecture slides, and *students support*, i.e., all the images generated by the teacher any time a student needed help in generating their images during the workshop.

On the teacher side, the images were collected by downloading these from the personal archive of the teacher on the *Midjourney* platform and counting them. On the students' side, the collection was more laborious. As we did not have access to the students' personal archives, we downloaded the entire chat of the Discord channels used during the workshop to interact with the *Midjourney* bot. The list of queries was downloaded as an excel file using a Chrome extension called *Discordmate*, which allows for bulk download of Discord chat content. After the bulk download, we cleaned the data to remove the queries from the teacher and the queries in which students forgot the command */imagine* (and thus generated no result). After the 'cleaning', the count of pictures needed a further step as queries included both groups of 4 pictures (Midjourney initially always generates a composition of 4 alternatives) and single images selected from the groups. To identify the single images in the list of queries, we searched for prompts containing *image #* corresponding to single images selected among the groups of 4, and prompts containing *variation*, corresponding to single images generated starting from previously selected images.

### 4.2. Energy costs of a two-day workshop on genAI for designers

We estimated the energy costs of the workshop by multiplying the amount of generated images by the energy consumption required by generating one image —0.0029 kWh— which we extracted from a recent estimation of energy costs of generating 1000 images (2.907 kWh) by Luccioni et al. (2024b). In total, the workshop generated 11072 images, using 32.1kWh, which is more than five times the daily energy consumption of a two-person apartment[1]. Beyond looking at the totals, we also calculated the *conversion rate* between the number of generated images and images that were used (both for teacher and students), and the *average individual energy consumption* (for students).

#### 4.2.1. Teacher

The teacher has generated a total number of 602 pictures over 10 days, including 554 pictures generated during the preparation of the workshop, and 48 generated during the workshop itself, to support students' work. Of the 554 images generated during the preparation of the workshop, only 41 (7%) made it to the lecture slides. Multiplying the total amount by the estimated electricity consumption of 1 generated image, we estimate that the activities of the teacher required 1.74 kWh.

#### 4.2.2. Students

The 49 students generated a total of 10470 images over the two-day workshop. Of these, only 512 (4.8%) made it to the *FigJam* board. Hence, the total activities of the students required around 30.36 kWh with an average of 0.6 kWh per student.

### 4.3. Grasping the scale of impact

To better grasp the scale of impact that running genAI educational activities has in terms of energy costs, here we first contextualize the results within related practices (Table 1) and then conceptually position our activity within the broader landscape of design education practices that are changing in response to AI.

| Unit | kWh | Source |
| --- | --- | --- |
| 1 hour of genAI workshop per student | 0,05 | Based on the average energy consumption per student we calculated based on (Luccioni et al., 2024b) (0.6 kWh) divided by the number of workshop hours (12) |
| Mid-range image rendering for 1 hour | 0,07 | Value estimated by (Iatan, 2017) |
| Laptop use for 1 hour | 0,054 | Based on the values reported by (Osthoff & Deakin, 2021) about a 2019 M1 MacbookPro, that is considered a highly energy efficient laptop and top choice for creative professionals |

**Table 1. Average energy cost of the genAI workshop for 1 student per hour, expressed in kWh, compared to the energy costs of related energy-intensive practices: image rendering, and laptop use.**

#### 4.3.1. Contextualizing data within related practices

The hourly energy consumption per individual student participating in our workshop (0.05 kWh) might look small at first glance –it is roughly in line with laptop use, and slightly lower than rendering high quality images for production. The comparison with image rendering, however, is only partially valid because of the different nature of the images' use within the practice of design. Currently, AI-generated images are mostly used in design processes to support ideation (Koch et al., 2019; Tholander and Jonsson, 2023; Chiou et al., 2023), provide inspiration (Kalving et al., 2024; Liu et al., 2024a; Ranscombe et al., 2024; Ling et al., 2024), and enable quick communication with clients and among team members (Zhao, 2024; Verheijden and Funk, 2023). After these phases, final project productions usually still necessitate professional image productions: the energy costs associated with AI-generated images are *in addition* to the existing costs such as producing professional image renders. As an illustration, Patrik Schumacher, studio principal at Zaha Hadid Architects, in a recent talk explained that the studio largely uses genAI for early ideation and selected images, and how about 10% of the total gets then moved to the 3D modelling phase (Schumacher, 2023). This underscores that genAI workflows are not a substitute for existing creative practices, but an addition to the process.

---

[1] The energy consumption of a two-person apartment is about 6 kWh per day, based on the yearly estimation by (SvizzeraEnergia, 2021) of 2190 kWh per year, and the one by (IRSAP2023) of 2000-2700 kWh per year, which are based on the average energy consumption of a series of appliances commonly found in two-person apartments, that are: cooking, dishwashing, and refrigerating appliances, lightning, clothes washing and drying appliances, entertainment electronics, small appliances, and electricity structural appliances, e.g., for heating.

The comparison with laptop use, instead, is especially important. Laptop use can represent up to 18% of a university's energy consumption (US Energy Information Association, cited in (Favela, 2023). Universities are large and resource-hungry institutions, with electricity consumptions that are often much higher than commercial offices (Munaro & John, 2024). They are pressured by growing demand for energy to support the Information and Communication Technologies they use (Sadorsky, 2012) on one hand, and sustainability goals such as the European Union directive (2023) to reduce end-user energy consumption by 11.7% by 2030 on the other. Because of the complexity of achieving these targets, scholars are starting to consider principles like energy sufficiency (Bodelier et al., 2024; Erba & Pagliano, 2021; Spangenberg & Lorek, 2019; Darby & Fawcett, 2018; Steinberger & Roberts, 2010) and suggesting policies that could set a limit on energy consumption per person (Faure et al., 2022; Spangenberg & Lorek, 2019) in addition to initiatives that use taxation as a strategy to reduce excessive energy consumption (Bertoldi, 2022). Accordingly, most universities have explicit plans for reducing their environmental impact and energy consumption specifically. This is exemplified by dedicated pages on universities' websites which illustrate the problem of energy consumption related to campus activities and, among other initiatives, emphasize the importance of exercising individual responsibility (see (Northwestern, 2024; Edinburgh 2024; Oxford, 2023; Uppsala, 2022; Unibo, 2020) for some examples). These usually encourage a conscious approach to the use of computers, suggesting simple actions for university populations to adopt, such as adjusting power and sleep settings, unplugging chargers, reducing screen brightness, and reducing image quality during online meetings, just to name a few. Along with these policies and recommendations, some institutions even developed initiatives dedicated to sensitizing students about the issue, providing 'energy literacy', and ultimately reducing the universities' computing impact (GEF, 2023; Saputra, 2022). These initiatives often take the form of *computing auditing actions*, to be performed in class or computer labs, that provide students and staff with a structure to make a record of all computing devices, i.e., desktop computers, laptops, printers, etc., and calculate the energy costs of these for a day (GEF, 2023).

In the face of this problematic reality and the strong effort towards *energy conservation* (Allen & Marquart-Pyatt, 2018) that universities face, educational activities involving the use of genAI induce students towards practices that move in the opposite direction. As shown in Table 1, *running genAI activities focused on image generation in design education can easily double the energy consumption of laptop use*.

### 4.3.2. Positioning the workshop within the landscape of design &AI education.

The case we took as an example in this work is only one among the multitude of AI-related educational experiences that the authors have carried out in the last year. The first author alone has run seven educational programs similar to the one we used for the calculations, and the second and last authors share a similar experience. Thereafter, we argue that to better grasp the scale of the impact associated with this type of activity, one should also consider how this is positioned within the broader landscape of design education engaging with genAI.

Our recent activities reflect a trend that is not peculiar to our personal educational experience, but rather part of a phenomenon that is affecting the design education field more broadly. Accordingly, we suggest considering our data as a single instance in a constellation of emerging educational practices. A quick scan of academic publications shows that there is an increasing number of educational design programs (not to mention programs in other fields) that are integrating genAI and being run across the globe: from weeks or semester-long courses (Kharrufa & Johnson, 2024; Li et al., 2024; Murray-Rust et al., 2023; Ching et al., 2023; Rajabi & Chris Kerslake, 2024; Simeone et al., 2022), to workshops and sessions lasting days or hours (Ali et al., 2024; Lee et al., 2024; Ray & Tang, 2023; Williams et al., 2024; Theophilou et al., 2023; Goel et al., 2023; Muji et al., 2023; Mariescu-Istodor & Jormanainen, 2019; Fiebrink, 2019). We are increasingly seeing genAI used in classroom settings with tens or even hundreds of students. Furthermore, these activities add up to the existing and constantly growing students' self-initiated use of genAI higher education. As found by a recent study conducted in Germany (Von Garrel & Mayer, 2023), more than 60% of students already use genAI tools, such as ChatGPT, in their studies. In short, design education activities engaging with genAI not only can double the energy expenditure we already have for individual laptop use but also come embedded within a global trend that moves in the opposite direction of societal efforts to reach sustainability goals.

## 5. ARTICULATING THE SPACE OF ENVIRONMENTALLY CONSCIOUS DESIGN EDUCATION AROUND GENAI

Our calculations around the environmental impacts of the workshop were prompted by a deep engagement with the current literature from both computer scientists and philosophers of technology, warning us about the dramatic environmental impact of AI. Perhaps influenced by classical mental models we hold about unsustainability (Cloud, 2014), in our discussions we engaged in a series of defensive positions. We scrutinized the argument that *responsibility lies with major players*, a view often strengthened by media headlines (see the New York Times one that states "*Tech giants are building power-hungry data centers to run their artificial intelligence tools*" (Sorkin et al., 2024)), which often point to tech giants as the only responsible actors. We also considered the positive possibility that, as with many other technologies, *AI algorithms will become more efficient* as the field becomes more mature. As the suggested by Schwartz et al. (2020), the field of AI should and can move from *Red AI*, with environmentally unfriendly and prohibitively expensive models, to *Green AI* where addressing efficiency is a primary evaluation criterion alongside accuracy. And finally, we extensively discussed how *there are already plenty of activities that consume a lot of energy that are part of our everyday lives* such as gaming (Mills & Mills, 2016; Perez et al., 2024) and video streaming services (Shehabi et al., 2014; Afzal et al., 2024; Gnanasekaran et al., 2021; Hossfeld et al., 2023) and whether the use of genAI should be regarded differently.

Although valid, these defensive positions fall short in the face of the impact these activities have on the already high environmental costs of higher educational institutions. While universities, and society at large, struggle to achieve sustainability goals, and individuals experience continuous pressure to reduce their energy consumption, *we are developing educational programs that quickly double the energy costs associated with students' use of computers*. Thus, we believe that as a design education community, we hold a responsibility to be aware of the problem and share this awareness with students too, through educational formats that account for genAI impact. Yet, as educators we also find ourselves overwhelmed by the need to reinvent educational practices to nurture AI literacy in future designers (Chan and Colloton, 2024) while being asked to do that without causing harm to our environment (Holmes et al., 2023). Furthermore, despite our identifying as critical scholars sensitive to matters of environmental justice, in our teaching experiences, we also found ourselves acknowledging the strong potential for engagement and exploration that these technologies have on students, and the transformational power that AI possibilities have on design processes. Over the last few years, we have observed how intrinsically motivated students are when designing with genAI, which contrary to some common beliefs (Peng et al., 2024; Subramonyam et al., 2023), is all but trivial. Through these experiences, we also learned how different the challenges and potentials of using these technologies are, whether we are teaching design for speculation compared to traditional product development (just to make one example). There is an argument to be made about the value of engaging directly with the materiality and dynamics of genAI tools (Lupetti & Murray-Rust, 2024). But a question remains: *is it worth it?*

A comprehensive, universal answer to this question falls outside the scope of this paper. Instead, we want to encourage the design education community to engage with it and make responsible individual choices. In support of this, we suggest *five alternative stances* one can adopt towards AI environmental impacts in education, along with actions consistent with each stance. These have been distilled through our process of critical reflection, which took the form of online discussion sessions. We started from the experience and the data described in the previous sections (3 and 4) to build self-awareness about our own stances and identify possible actions to practice sustainability in design education when engaging with genAI. These are not intended as turnkey solutions, but rather as a source of inspiration and starting points for guiding personal reflections of design educators finding themselves navigating similar concerns.

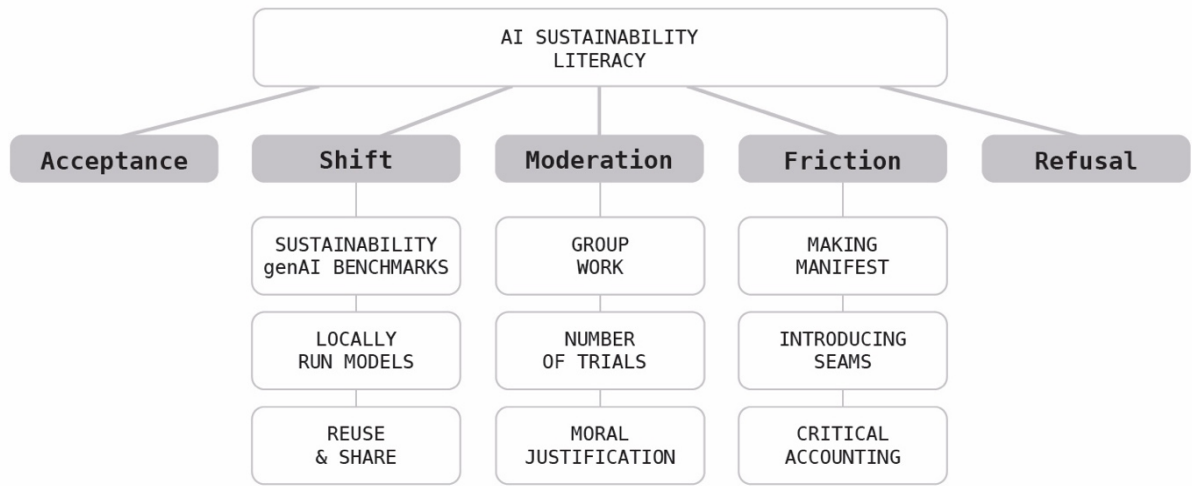

**Figure 3. Building blocks of Sustainable AI practices in design education: possible stances (grey) and related activities (white). Stances are detailed in their own sections: Acceptance (5.2), Shift (5.3), Moderation (5.4), Friction (5.5) and Refusal (5.6), and the overarching layer of AI Sustainability Literacy is in Section 5.1.**

### 5.1. AI Sustainability Literacy

Developing literacy about the sustainability problem surrounding genAI cuts across all of the stances that we set up, and is a required foundation for meaningfully addressing the responsible use of genAI in design education. While significant attention has been given to the social and technical issues of AI uses, given the climate crisis we live in (United Nations, 2020) and the tremendous global impact that recent genAI developments have had (Bashir et al., 2024), there is an urgent need to increase understanding of the materiality and environmental costs of genAI, bringing this in line with growing socio-technical literacy. There are several technical resources that can be used to nurture this understanding, such as the work by Luccioni et al. (2024) on AI electricity costs, by Dodge et al. (2022) on carbon intensity of different AI models, or by Li et al. (2023) on AI consumption of clean freshwater. Alongside these technical resources, an increasing amount of design-friendly materials is becoming available, such as critical cartographies (Joler, 2024) that map out the systemic nature of these systems, such as Crawford and Joler's Anatomy of an AI System (Crawford & Joler, 2028), or the Cartography of GenAI by Estampa (2024), or detailed reflections on AI's effect on the planet (Brevini, 2022).

### 5.2. Acceptance

*"It is important to be aware of and talk about the costs of genAI, but AI literacy and competence has priority"*

At one end of the spectrum, the position of acceptance is one in which AI sustainability concerns are tolerated - accepted - for the sake of nurturing students' AI literacy. Here, equipping students with knowledge and capabilities for skillfully engaging with AI is seen as an urgent need, of greater importance than sustainability. To grasp this position, a parallel may be made with the environmental costs of training jet fighter pilots that are extensive, yet acceptable because of the critical role they play in case of need (McCarthy, 2019). This may be judged as a relentless or selfish choice, yet we encourage looking at it as a position coherent with the very scope of design education institutions. Indeed, many educators do not engage with the problem of genAI sustainability and instead focus on socio-technical aspects of AI education, at least based on the state-of-the-art literature on AI literacy for designers. Educators may confront themselves with the problem of environmental impacts, and still decide to integrate genAI in their teaching practice. This position is in line with the view of Maceli and colleagues (Maceli et al., 2024) who argue that we should not prohibit students from using genAI tools, but rather we should focus on nurturing a critical understanding of these technologies so that they will be enabled to make conscious choices in the future. Accepting the costs of genAI and deciding to engage with it in education, however, does not necessarily mean neglecting the sustainability problem. In this position, the aspect of environmental sustainability can be addressed as one of the factors contributing to the eco-socio-technical complexity of AI system that students should be literate about (Rakova and Dobbe, 2023).

### 5.3. Shift

*"It is important to minimize the impact of genAI teaching activities by making more sustainable choices about the technologies that we use. This will help students afterwards to make better choices in their practices"*

The position of shift comes from a similar standpoint to acceptance, i.e., that the need for AI literacy is of higher priority in education than the one of minimizing environmental impact. However, it brings a distinct set of practices. This is conceptually aligned with the technical and regulatory approach of designing for sufficiency (Bodelier et al., 2024; Darby & Fawcett, 2018; Erba & Pagliano, 2021; Spangenberg & Lorek, 2019; Steinberger & Roberts, 2010) where educators actively seek to minimize their impact by shifting toward 'better' genAI choices. We identify three main types of practices in this space.

#### 5.3.1. Sustainability genAI benchmarks

One way to exercise a responsible engagement with genAI in education is to carefully select models and tools that are smaller and more efficient. In order to make informed choices in this direction, it is of crucial importance to develop and/or use benchmarks through which one can compare models' performance with 'costs'. Benchmarks do exist and most common examples focus on the ratio between performance and financial costs, i.e., submission prices or price per token. For instance, more and more benchmarks show how tiny models, such as GPT-4o Mini, Mistral NeMo, and Llama-3.1, actually achieve high performances while relying on smaller amounts of data, and thus being more efficient (OpenAI, 2024; Heka AI, 2024). Resources that account for efficiency and environmental impact are starting to emerge yet remain limited and difficult to grasp for non-experts (see (Asperti et al., 2021; Chen et al., 2023; Yu et al., 2022). To our knowledge, the most detailed yet accessible sustainability genAI benchmark available today is the one by Luccioni et al. (2024) who tested 88 models, diverse in terms of purposes and characteristics. However, making these resources usable and meaningful for design educators requires continuous updating, as novel models or versions of existing models are constantly being released.

#### 5.3.2. Locally run models

A related but distinct practice is to encourage the use of genAI models running on local servers. Not relying on cloud computing, locally run models allows for reducing reliance on data centers and, thus energy consumption. We see moves in this direction, such as "we will outline some tools that enable offline image and text generation and provide links to their quickstart guides." (NYU Libraries, 2024). There are several advantages to running local models. Firstly, these models tend to be smaller in the number of parameters and more efficient, hence likely to have reduced energy costs. This can be seen with models like Llama 8B, which is optimized to provide a high-quality LLM that can run on a laptop, or Gemini Nano (Anil et al., 2023) which targets devices such as high-end mobile phones. Secondly, working with local models brings the compute closer to home: it becomes more obvious that the computation is power intensive when the machine starts to heat up, the fans engage and other operations become slower. Users of local models may also have to make tradeoffs in terms of space - at 8GB, a language model uses a significant amount of disk space for a laptop, which helps to push towards finding smaller solutions. Finally, using local models alters what the upper bounds on usage are: where a cloud-based model is typically bounded by either tokens or usage rates, a local model is bounded by the compute – and power – available to the machine being used.

#### 5.3.3. Reuse and share

Another way to reduce impact is to choose optimized resources. This can apply to both the training and the querying of models, but adopting a reuse & share approach to the first is particularly impactful. When looking at LLMs or image generation, the cost of training the models is extremely high, but the dominating cost tends to be the inference, both because it is costly, and because it happens at scale. As an example, training a RAVE model (Caillon & Esling, 2021) for music or audio takes approximately three weeks of a high-powered GPU running full-time – a relatively high cost for developing an individual model. However, the trained model can run inferences in under 200ms on a modern laptop – relatively, a much lower resource cost. This is a space where people tend to train their own models, seeing it as part of the creative process. However, communities that are happy to share trained models – such as the Intelligent Instruments Labs collection of RAVE models based on sonically interesting training data (Intelligent Instruments Lab, 2023) – allow the costs of training to be spread across a far higher amount of creative practice through creative re-use. As custom-trained models become increasingly part of practice, finding good ways to share and re-use others' work, while still developing an understanding of the connections between models and training data, can be part of efficient model use.

## 5.4. Moderation

*"It is important to engage with genAI in education, but its impact must be moderated by reducing the number of operations or requiring a moral justification for each use"*

Similar to shift, the position of moderation is close to the technical and regulatory approach of sufficiency (Bodelier et al., 2024; Darby & Fawcett, 2018; Erba & Pagliano, 2021; Spangenberg & Lorek, 2019; Steinberger & Roberts, 2010. Distinctively, however, this concept inherently comes with an element of limitation not on the technology itself, but rather on the activities using it. The is an implicit imposition of boundaries to the use of genAI in educational practices. Here we illustrate three ways to practice moderation.

### 5.4.1. Group work

One can prioritize group work over individual exploration. This can be done by providing a single license for each group (for genAI tools that require one like MidJourney), which could greatly reduce the amount of generated images and thus, the energy consumed, because of the impossibility of simultaneous use of the tools. There is, however, an educational drawback here to consider. While prioritizing group work can encourage discussion among team members (Han et al., 2021) about the dynamics and results obtained from the genAI tools, it may also inadvertently encourage teams to distribute tasks, with the ultimate result of having a single person in charge of engaging with the genAI tools. Although this risk is highly dependent on the way educators introduce genAI in the design education program, the very activity with genAI may run into the risk of becoming a disjunctive task [38] that may further increase the technological competence divide between students, ultimately conflicting with the scope of nurturing AI literacy for all future designers.

### 5.4.2. Number of trials

A different way to moderate the use of genAI is to limit the number of trials one can use when prompting. This is opposite to the usual approach in which educators tend to let students try as many times as they want (as in (Lee et al., 2024)), which can be seen as a form of craft education, where genAI is approached as novel material subjected to experimentation, investigation, and invention (Niiranen & Rissanen, 2017), and the very deep engagement with the materiality of AI allows to meaningfully engage with its potential and limitations (Lupetti and Murray-Rust, 2024). Setting a limit to the number of times students can try and generate results with the genAI tools, then, may also conflict with the education scope of building AI literacy, but this risk may be mitigated by the provision of clear and detailed guidance by the teacher. Studies have shown how giving clear guidance on how to design a prompt, especially by providing a prompting manual, can have a very positive effect on students' user experience of genAI tools, i.e., in terms of enjoyability and ease of use (Lee et al., 2024; Theophilou et al., 2023).

### 5.4.3. Moral justification

The third, and last, way we identified for moderating the use of genAI in education is to bind it to design activities engaged with topics in which genAI may play a crucial role for societal good. An example of a morally justifiable application is the case of genAI being used in Mali to create, in less than a year, a collection of 107 illustrated schoolbooks for children in Bambara language, one of the local languages being taken up again after the military government sanctioned the abolition of French – colonial language (Risenberg & Dosunmu. 2024). Here genAI may be morally justified as the alternative may be limited access to education for children, at least for a certain period. Defining when and how genAI can be beneficial for society and thus morally justifiable, however, is non-trivial. Also, cases like the one of education in Mali could be approached in alternative ways. We see this not as a limitation, but rather as a potential strength for educational activities around AI, where students are invited to engage with questions that are central to the AI ethics debate, such as what makes the use of genAI morally justifiable and who gets to decide (Cavalcante-Siebert et al., 2023. Setting contextual boundaries, then, can be limiting in terms of possible creative uses of genAI allowed in the classroom, yet it would carry great value in fostering students' critical thinking around AI technologies.

## 5.5. Friction

*"Using genAI is important in education, but students must be continually confronted with the costs incurred"*

One of the issues with genAI, particularly in its cloud-based form, is the invisibility of the dislocated environmental costs. Images appear without a trace of their construction, so there is no connection to the material workings of the data centre that produces them. Engaging with friction is often part of developing a critical technical practice (Agre, 2014; Hirsbrunner et al., 2024) that develops reflexivity about the use of technical artefacts and systems. Creating sites for co-reflection, discussion and sharpening of positions helps to negotiate shared boundaries (Bertelsen & Bødker, 2002) as students unpack the ways that they relate to the hidden infrastructures and unpack their values (Forlano & Mathew, 2017). Speculative frictions are already being employed to think about AI and data governance (e.g. https://speculativefriction.org/ (Rakova, 2019)), and we can expand these practices to challenge questions of sustainability and environmental impacts in the classroom.

### 5.5.1. Making Manifest

Educators can invite students to engage in projects that manifest the costs associated with their activities. For this, one can leverage methods and tactics consolidated in the space of critical design, in which several examples can be found of how to manifest various forms of energy cost and practices around them. For instance, Bodelier and colleagues (2024) developed a speculative router that visualizes the amount of internet data being used by different people in the same household, as a way to encourage self-awareness and social pressure to reduce personal data consumption. Other examples include Lindley's *Cryptoheater* (Lindley, 2015) that manifested blockchain computation as heat, or Kyle McDonald's "Amends" which counts the carbon cost of the Ethereum system used to create online art marketplaces, and offers the sculptures for sale for the cost of the necessary carbon offsets (McDonald, 2022). Particularly in an educational setting, this manifestation of costs can have an impact - it connects actions to their effects, so rather than refusing to use the technology, educators start to enable meaningful comparisons to be made by the students, as part of a sustainable design practice.

### 5.5.2. Introducing Seams

Alongside manifesting costs overall, introducing some seamfullness or barrier at the moment of use can respond to the otherwise seamless nature of engaging with genAI (Bødker, 2011). In a typical educational journey with technology, there is an initial hurdle to getting the technology working: setting up development environments, installing the right software, developing a conceptual and practical understanding of how the components fit together and so on. Normally, the hope is that this technological burden disappears - and in practice, we often see that after this initial push into the unknown, students dive into rapidly exploring the possibilities of their new abilities. One easy way to introduce an element of friction here would be to shift the stance from 'aiming to make sure there is enough credit on the OpenAI account for the students to run the examples' to 'purposefully putting a small amount of credit on, and then make top-ups explicit'. This easily broadens the educational stance from wanting the smoothest possible experience to valuing the seam (see also Making Visible in Blockchain and Beyond (Murray-Rust et al., 2023)).

### 5.5.3. Critical Accounting

Accounting for the costs of creating AI-generated artefacts in education is a powerful way to help develop a critical mindset. This is somewhat related to moral justification mentioned above, where justification is a blanket license that 'this kind of thing is OK', critique could involve the friction of counting and accounting for each use, with the act of 'keeping score' a key part of the technical practice (in a similar fashion to the work we did in this paper). Similar techniques are appearing more generally: asking students to explain the details in an AI-generated image quickly illustrates that while the technology makes any image visually 'shiny', it doesn't automatically flesh out an underdeveloped concept or underspecified instructions in a meaningful way. This practice is also conceptually aligned with the recommendations of the European Union (2024) for industries to report and document their AI development processes in terms of resource performance and energy consumption. Accounting for the creative decisions made is then a way to introduce criticality about the roles of students, and in a similar way accounting for the costs of AI uses friction to contextualize the practices. Beyond counting images generated, as resources for calculating impacts become available—for example the user-facing ML calculator (Lacoste et al., 2019a; Lacoste et al., 2019b), or the more programmatic MLCarbon (Kaneda, 2024), educators can ask for a similar accounting from students. This kind of accounting can form the basis of a reflexive section in the assignment report, in line with the ways that students are required to sustainability assessments of their designs, and commercial products report on proportions of recycled materials or the repairability of manufactured devices.

### 5.6. Refusal

*Even though genAI is important, the costs of using it in education outweigh the benefits*

Some educators may feel that the environmental costs of using AI are not justified, either in the courses that they are teaching or more generally in any case. Rejection of genAI can arise from various concerns beyond sustainability. The possibility of refusing or limiting the use of genAI in classrooms emerges particularly when it is believed that these tools might hinder the development of essential skills, especially for designers, such as the ability to generate original ideas. The question is often posed in absolute terms: "Should we prohibit the use of generative AI?" highlighting a critical stance that tends to preclude a balanced discussion. Here we would separate the refusal on environmental grounds from others based on pedagogical concerns, such as the ones about plagiarism, authorship, untrustworthiness or questions of what parts of the educational experience students are missing out on (Licht, 2024; Kishore et al., 2023). Of course, in practice, such refusals are often interconnected, and may not be separately articulated either internally or externally by the educators; here we focus on the environmental aspects of refusal. Beyond simply not using genAI, educators can make their position clear as part of their activities. This can be simple statements about their value systems or more detailed explanations of their cost/benefit analysis around AI.

## 6. DISCUSSION AND CONCLUSIONS

In Sections 5.1 to 5.5, we articulated a range of stances that design educators might take around the use of genAI in their programs. These stances are not exclusive but rather highlight points in a landscape of possible relations to using genAI in education. Except for acceptance and refusal, which are the radical ends of this positionality space, standing somewhere in between allows for mingling with all stances. One can in fact develop educational programs that implement several actions we associate with different stances, and most likely, this is what realistically happens. For instance, the first author has, on other occasions, run workshops that deliberately combined the use of predefined prompts (reuse & share) to be personalized by students, along with group work. In others, an image prompting activity was turned into a tangible interface first (introducing seams), so that students could familiarize themselves with a prompt structure, and then performed on a computer later, limiting the number of trials each student could do. While not exhaustive and non-exclusive, these stances and related practices are seeds for conscious educational practices. These simultaneously activate the need for balanced thinking and trade offs, and encourage to make the difficulty of counting the impacts part of the pedagogical practice, while giving space to students for making use of the emerging tools. Further, the stances give educators a vocabulary to describe the positions they take. By nuancing the space between acceptance and refusal, these can support a combination of discourse and practical action, that together can create long-term changes. In what follows, we further discuss the complexity of practicing counting activities like the one presented in this paper, and the pressing need for embedding accounting practices in the design education more broadly.

### 6.1. A blueprint for counting

As previously argued, a hope for this work is to encourage the design education community to engage with the genAI sustainability problem and make responsible individual choices. For this, it is crucial to build practices that allow for personal quantification of impact as a basis for further actions. We illustrate in Section 4.1 the process used to perform calculations ourselves, about a particular workshop. This, we believe, can serve as a *blueprint for counting*– a resource for others to structure similar impact estimation activities. In the same vein as our calculations, educators can run a program, afterwards collect and clean the data, and then proceed with the calculations based on the values by (Luccioni et al., 2024b). Putting this blueprint into action, however, is non-trivial and the details of the final procedure may need adaptations, depending on a variety of factors.

First of all, there is a facilitating element in the process we applied, which is the use of *MidJourney* instead of a multipurpose genAI platform. This is specialized in producing images and is peculiar in the fact of leveraging

*Discord* as an interface platform, as opposed to the vast majority of genAI products that run on dedicated platforms. Because of that, collecting data (both educators' data and students' data) after activities is extremely easy. Other image-specific genAI tools, such as *Leonardo AI*, are to some extent similar and allow for easy counting of generated images. Conversely, data collection on some genAI platforms may be more laborious, up to becoming not feasible. If using ChatGPT, for instance, one can save a chat history by downloading personal ChatGPT data (OpenAI, 2024), then open the data in Excel, select the relevant chat and finally search for "DALL-E" (the image model used by ChatGPT) mentions within the chat. This is more laborious compared to our process, yet doable within a limited time and with a relatively small effort. In other tools, like *Gemini* and *Co-Pilot*, instead, it is unclear whether it is possible to download a chat history, leaving direct searching and counting in the chat itself as the only possibility for counting. This may be doable within a generative session on a chat that is not long (i.e., from a short class or workshop), but it becomes impossible for counting generated images in case of workshops lasting days, or semesters. This is further complicated by the often hybrid workflows that students (and educators) carry out. These use multiple processes with varied impacts and often present a back-and-forth between different tools, including non-AI tools (i.e., Photoshop, Illustrator and more). Furthermore, hybrid workflows are implicitly encouraged by the spread and use of "AI aggregator platforms", such as *Poe*, where students can access and use a variety of different models with a single account. Furthermore, some tools not only come with different –lower– ease of accessing and processing data but also allow only for collection from the individual user, making it impossible for an educator to conduct counting activities for an entire class.

Finally, keeping track of genAI use has limitations that come with the model benchmarks we use for running the calculations. While figures provided by Luccioni et al. (2024b) give a place to start, these may soon become outdated as AI models are continually evolving, both in terms of efficiency and ambition, so without detailed information from vendors, precise estimates are challenging. Also, here we have provided an example of accounting focused on image counting, which is a particularly practical estimation activity. While estimating the energy costs of generated text in design education projects may also be feasible, it is hardly possible for other media, such as AI-generated videos or text-to-speech audio files. For these, we simply lack benchmarks.

These practical challenges, however, shall not be seen solely as limitations, but rather as *an encouragement for defining novel educational practices*. Assignments can incorporate first-person counting of generated images which, on the one hand, allows educators to estimate the impact of their planned activities and adjust accordingly, also in light of the stands and actions we articulated in 5, and, on the other, allows students to estimate the impact of their individual activities and reflect upon that. This can integrate existing reflexive practices that already are integrated into the assignment reports that students have to submit at the end of a course, in the form of a reflection section (see for instance (Murray-Rust et al., 2023)). Further, developing the *AI sustainability benchmarks* we need as educators can set an agenda for researchers from the design field and beyond.

### 6.2. A call for accounting

By encouraging to count and consider environmental impact as integral to educational programs, this work invites the design education community to develop novel practices. Underlying our call for more environmentally conscious actions, however, a question remains: *do we have a sufficient culture of sustainability in our educational practices?*

It is easy to point fingers at the new technology being introduced – it is a heavily critiqued, potentially transformative, but environmentally costly way of working – yet it is important to contextualize this in the material histories of design education more broadly. In the 1960s, the introduction of plastics saw an explosion in the range and fluidity of forms and colors that could be created (Meikle, 1995), and a vast array of properties, such as limited weight and sterility, with an inevitable rise in the environmental impacts. Papanek's famous quote— "*There are professions more harmful than industrial design, but only a very few of them*" (Papanek, 1972, p.14)—illustrates the long-running issue of accounting for the harms of design practices, within and beyond education. Yet, to our knowledge, there has never been an educational culture of accounting for the environmental impact we have as a design educational institution, i.e., *has anybody ever counted the amount of Styrofoam that each individual student used in a classic design studio course where a physical product prototype was expected as deliverable?*

The sustainability issue has recently been gaining new momentum in design education. This can be seen in the work of the Future of Design Education group (Future of Design Education, 2024), and their recommendations on sustainable practices in design education (Faludi et al., 2023). Many of these resonate with the advice given here: developing literacies, bringing systems thinking into design practice, justifying work in terms of e.g. environmental justice and so on. Overall, the work presented here fits into a larger transition that design educators are having to make, to consider the sustainability of both immediate practices and longer-term implications of their design education. Furthermore, many of the actions we suggest here draw on this field and fit within the broader space of critical design practices, for which, however, the field might need new understandings and resources. Critical design practices have been accused of failing to address the problems that they condemn (Ward, 2021) and often not dealing with environmental issues. As a historical example, the Blow armchair by Zanotta (2024), designed by De Pas, D'Urbino and Lomazzi in 1967 was created as part of an emerging critical practice set in opposition to the modernist and rational status quo at the time. However, the issue of environmental sustainability was not part of such critique, and products like this contributed to what, with time, became a mainstream design practice with dramatic environmental costs, i.e., made of un-recyclable plastics, with a strong dependence on the world supply of oil (Nannini, 2023). While the critical design of the 1960s sought to challenge the status quo, it often neglected the ecological dimension, similar to what is happening now with many critical initiatives around AI that condemn socio-ethical issues but fail to address matters of environmental justice (Rakova and Dobbe, 2023). In the context of the growing climate crisis, however, this is not a 'lightness' we can afford anymore. But there are possibilities to move beyond this and seek "tension with progression", where the frictional tendencies of critical design (Pierce, 2021) create a space between reflection and production (Lupetti, 2022). Hence, the suggestions given here try to mediate, and inhabit the space of tensions, as a productive approach to dealing with the competing drivers of providing cutting-edge education in a changing world and creating climate-literate, sustainability-sensitive design practitioners, as well as bringing these questions into the design of curricula.

But even when aware of the issue, dilemmas might remain. One can argue that there is always some environmental impact when carrying out education, particularly around material practices, in which case we should ask: *how do we balance the educational need for iteration in design with the associated environmental costs?* Much of design education is concerned with the very doing of things, of experiential learning as a key part of developing skills and fluency (Dutton, 1987). The 'practice' in critical practice (Agre, 2014) implies multiple attempts, and a gradual honing of skills (Wright, 2011). As for the action of moral justification (Section 5.3.3), here we encourage the community to ask who decides and on what grounds, that the costs of genAI in education are justifiable (Cavalcante-Siebert et al., 2023). In doing so, the work also invites the community to reflect on the power and responsibilities we have as educators, as we also make decisions for tens –even hundreds– of students who might not share our personal stance on the matter of AI sustainability.

In short, this work calls on educators to critically evaluate the overall impact of their educational programs, integrating sustainability as a fundamental element in the design of their activities. Only through critical reflection and a commitment to more conscious practices can we hope to prepare future generations of designers not only to utilize innovative tools but also to do so in a sustainable and responsible manner.